\documentclass[useAMS,usenatbib]{mn2e}

\usepackage[dvips]{graphicx}
\usepackage{natbib}
\def\kms{$\rm \,km\,s^{-1}$}

\def\H2{H$_2$}

\def\micro{$\rm \,\umu m$}
\def\arcsec{\,arcsec}

\title[MIR imaging of NGC1068 by VISIR]{Mid-infrared imaging of NGC1068 with VISIR at the VLT}
\author[E. Galliano et al.]{E. Galliano$^{1}$\thanks{E-mail:egallian@eso.org}, E. Pantin$^{1,2}$, D. Alloin$^{1,2}$ and P.O. Lagage$^{2}$\\
$^{1}$ESO, Alonso de Cordova 3107, Vitacura, Casilla 19001, Santiago 19, Chile\\
$^{2}$UMR 7158, CEA - CNRS - Universit\'e Paris 7, DSM/DAPNIA/Service d'Astrophysique, CEA/Saclay, France}

\begin{document}

\date{Accepted xx. Received xx; in original form xx}

\pagerange{\pageref{firstpage}--\pageref{lastpage}} \pubyear{2002}

\maketitle

\label{firstpage}

\begin{abstract}
High resolution mid-infrared (MIR) images of the central region of NGC1068 have been obtained with VISIR, the multi-mode MIR instrument recently installed at the ESO/VLT on Paranal. A map of the emission at 12.8\micro~with increased sensitivity over the central 8$\times$8\arcsec$^2$~area is discussed. It shows a central core (unresolved along the E-W direction) and an extended emission which draws a spiral pattern similar to that observed on near-infrared images. Patches of MIR emission can be detected up to a distance of 4\arcsec~from the core. The deconvolved 12.8\micro~map is fully consistent with previous high-resolution MIR observations. It highlights the structure of the extended emission, already seen on the un-deconvolved image, and allows to identify a set of mid-infrared sources: 7 in the NE quadrant and 5 in the SW quadrant. The MIR emission map is compared with those obtained at comparable angular resolution in the near-infrared and in the [OIII] line emission. The very good correlation between the VISIR map and the HST optical map supports the idea that the MIR emission not associated with the torus arises from dust associated with the narrow line region clouds. The N-S extension of the MIR core (0.44\arcsec) is then probably simply due to the mixing of the MIR emission from the dusty torus and the MIR emission from NLR cloud B, located only 0.1\arcsec~to the North. 
\end{abstract}
\begin{keywords}
galaxies: active, nuclei -- individual: NGC1068 -- infrared:galaxies
\end{keywords}
\section{NGC1068, a privileged target for the VLT/VISIR Science Demonstration Program}
\label{introduction}
\begin{figure*}
\begin{center}
\includegraphics[width=18cm]{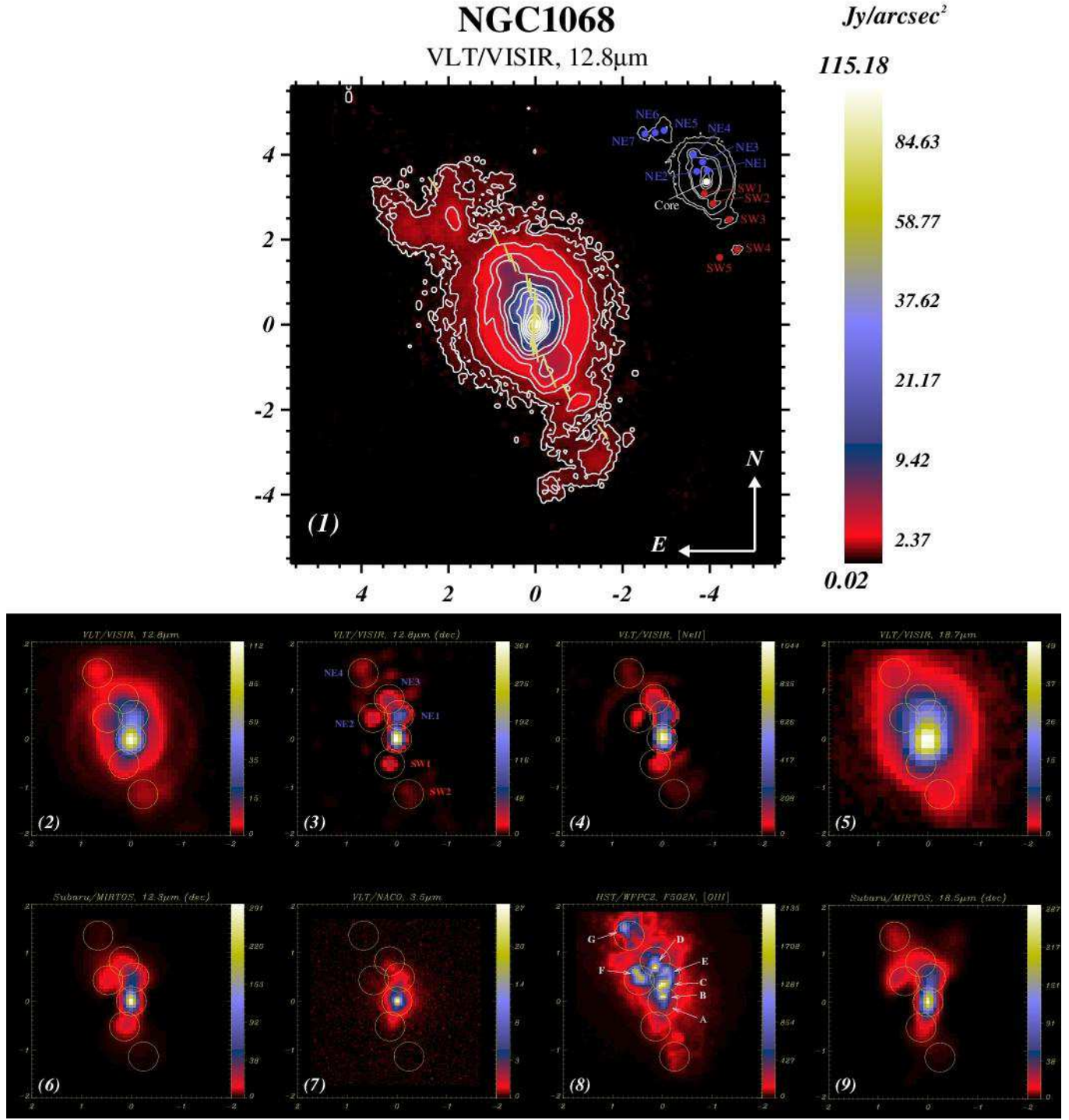}
\caption{For all the images: The X- and Y-axis are in arcsec, North is up and East is left, and the (0\arcsec;0\arcsec) position corresponds to the location of the central engine as derived in \citet{Galliano03b}, coincident with the MIR core. Images 1 to 5 were recorded by VISIR at the VLT. The colour scale represents either the continuum flux density surface brightness in Jy\,\arcsec$^{-2}$ (all images but 4 and 8) or the line flux surface brightness in 10$^{-14}$\,erg\,s$^{-1}$\,cm$^{-2}$\,\arcsec$^{-2}$ (images 4 and 8). The circles are drawn at the positions listed in table~\ref{table} and have a diameter of 0.6\arcsec.   
\textit{(1)}: 12.8\micro~un-deconvolved image of the central region of NGC1068 by VISIR. Contour levels are 0.07, 0.08, 0.13, 0.30, 0.76, 1.7, 4.4, 7.2, 13.1, 23.1, 35.5, 55.1\,Jy\arcsec$^{-2}$. The yellow tick-marks represent the position angles measured through ellipse fitting at the distance from the peak corresponding to the centre of the segment. In the top-right insert, a sketch of the image has been drawn identifying and naming the MIR emission sources referred to in the text. \textit{(2)} central part of the 12.8\micro~image shown in (1). \textit{(3)}: deconvolved 12.8\micro~image. \textit{(4)}: deconvolved continuum subtracted [NeII] image. \textit{(5)}: 18.7\micro~filter image. \textit{(6)} Subaru/MIRTOS 12.3\micro~image \citep{Tomono01}. \textit{(7)} VLT/NACO 3.5\micro~image by \citet{Rouan04}. \textit{(8)} HST/WFPC2 F502N filter ([OIII] 5007\AA) image \citep[cloud names after ][]{Evans91}. \textit{(9)} Subaru/MIRTOS 18.5\micro~image \citep{Tomono01}.}  
\label{Fig1}
\end{center}
\end{figure*}

Thanks to its brightness and proximity (14.4\,Mpc; 1\arcsec=70\,pc), the active galactic nucleus (AGN) in NGC1068 has been studied in more details than any other AGN. It has played a key-role in the emergence of the unified AGN model: Antonucci \&~Miller (1985) showed that the polarised optical spectrum of this type\,2 AGN was identical to that of a type\,1. To understand this discovery, the authors proposed that a type\,1 AGN was embedded within an optically thick dusty torus, hiding the central engine and the broad line region (BLR) from direct view along specific lines of sight. In parallel, the dusty torus was invoked to explain the type-dependent infrared (IR) bump observed in the spectral energy distribution of AGN and the conical shape of the Narrow Line Region (NLR). The work of modellers \citep{Krolik86,Pier93,Granato94,Efstathiou95,Granato97} gave support to this interpretation and a unified AGN picture involving a dusty torus became commonly accepted in the community. 

Still, after several years of search, the existence of the torus has not been established in a decisive and direct way. Therefore, NGC1068 remains a natural priority target for the latest generation of high spatial resolution IR instruments. The high quality adaptive optics images obtained with NACO at the VLT \citep{Rouan04} in the K, L and M bands (0.06\arcsec~resolution in K) did not significantly resolve the central core in the E-W direction (see Sect.~\ref{the core}). Recently, NGC1068 became the first extragalactic object to be observed with the VLTI \citep{Wittkowski04,Jaffe04}: for the first time, spatially resolved informations on the central emission core were obtained. However, the visibility measurements collected so far remain too scarce for a decisive interpretation and the existence of the torus in this AGN is still under discussion. 

Besides the core emission and the reality of the torus, the existence of an IR extended emission region around the central engine is known since the works of \citet{Bock98}, \citet{Alloin00}, \citet{Bock00} and \citet{Tomono01} in the MIR, and \citet{Rouan98}, \citet{Marco00}, \citet{Thompson01}, \citet{Rouan04} and \citet{Gratadour05} in the NIR. On a large scale (up to 5\arcsec), the emission is surprisingly coincident with the radio jet observed by \citet{Gallimore96c}. At smaller scale (0.5\arcsec), it appears to be coincident with the NLR cloudy structure observed by HST \citep{Macchetto94}, using the relative positioning discussed by \citet{Galliano03b}. The physical processes at the origin of the extended MIR emission still need to be clarified. Even if the fact that silicate absorption is detected all over the MIR emitting region \citep{Tomono01,Galliano03b} supports an interpretation in terms of warm/hot dust, the coincidence of the MIR emission with the radio emission, on large scale, is not fully understood.   

VISIR, the new MIR instrument at the VLT, received first light in May 2004. NGC1068 was an obvious target for the Science Verification Program. Preliminary results based on these observations are presented in this Letter.

\section{Observations and data reduction}

\label{Observations}
VISIR\footnote{http://www.eso.org/instruments/visir/} is the recently commissioned MIR instrument installed at the Cassegrain focus of Melipal at the VLT. It is composed of an imager and a long-slit spectrometer. In its imaging mode, a large set of filters is available, covering both the N- and Q-bands. The possible pixel scales are 0.075\arcsec~pixel$^{-1}$ and 0.127\arcsec~pixel$^{-1}$ respectively. These sizes are small enough for efficient deconvolution.

NGC1068 was observed between 2004 November 24 and November 30 through four filters: the 12.8\micro~filter ($\lambda$=12.81\micro, $\Delta\lambda$=0.21\micro), the reference filters for [NeII], 12.3\micro~($\lambda$=12.27\micro, $\Delta\lambda$=0.18\micro) and 13\micro~($\lambda$=13.04\micro, $\Delta\lambda$=0.22\micro), and the 18.7\micro~filter ($\lambda$=18.72\micro, $\Delta\lambda$=0.88\micro). For the observations, the nodding was set parallel to the chopping, with throws of 12.5\arcsec~in the N-S direction. The total integration time on source was 1500\,s. Standard stars from \citet{Cohen99} were observed  before and after each target exposure, for PSF determination and flux calibration. The weather conditions were good, and the images are close to the diffraction-limit. 

Dedicated \texttt{IDL} routines have been written for the reduction of the raw data. This basically consisted in shifting and stacking the individual frames, each frame corresponding to one chopping position. The precision of the flux calibration is limited by variations of the conversion factor due to changes in the transparency of the atmosphere. The final uncertainty on the flux of a high signal-to-noise source is of the order of 20\%. The deconvolution was performed using the multiscale maximum entropy algorithm by \citet{Starck02}. The subtraction of the continuum from the non-deconvolved 12.8\micro~image was performed as follows: a 12.8\micro~continuum image was interpolated from the 12.3\micro~and 13\micro~images and then subtracted from the 12.8\micro~image (all images matching the same resolution). For this procedure, the relative positioning of the maps was performed according to the NE heart-shaped structure, pointed out by \citet{Alloin00}. 

\section{The data}
\begin{table}
\caption[]{Positions and 12.8\micro~flux densities of the sources in the central region of NGC1068 (from the VISIR image)}
\begin{center}
\begin{tabular}{|l|cccc|} \hline 

source & $\Delta\alpha$ & $\Delta\delta$ & error & Flux \\
     &      [arcsec]    & [arcsec]       & [arcsec] & [Jy] \\ \hline
$^{nd}$     NE7  &  3.07 &  2.40 &  0.08 & 0.05$\pm$0.04$^*$\\
$^{nd}$     NE6 &  2.47 &  2.40 &  0.08 &  0.05$\pm$0.01$^*$\\
$^{nd}$     NE5 &  2.02 &  2.70 &  0.08 &  0.07$\pm$0.05$^*$\\
$^{nd}$     NE4 &  0.67 &  1.35 &  0.02 &  1.$\pm$0.35\\
$^{ d}$     NE3 &  0.15 &  0.82 &  0.02 &  3.8$\pm$1.1\\
$^{ d}$     NE2 &  0.47 &  0.43 &  0.02 &  1.45$\pm$0.3            \\
$^{ d}$     NE1 & -0.05 &  0.49 &  0.02 &  11.2$\pm$2\\
$^{ d}$     Core & 0.0 &  0.0 &  0.0 &  25$\pm$5\\
$^{ d}$     SW1 &  0.14 & -0.52 &  0.02 & 1.2$\pm$0.2 \\
$^{nd}$     SW2 & -0.24 & -1.14 &  0.02 & 0.42$\pm$0.15\\
$^{nd}$     SW3 & -1.12 & -1.94 &  0.08 & 0.08$\pm$0.2$^*$\\
$^{nd}$     SW4 & -1.51 & -3.38 &  0.08 & 0.04$\pm$0.03$^*$\\
$^{nd}$     SW5 & -0.38 & -4.05 &  0.08 & 0.23$\pm$0.1$^*$\\ \hline

\end{tabular}
\end{center}
\label{table}
The flux densities are measured through apertures of 0.6\arcsec~diameter, centred at the position given in the table. The errors take into account the uncertainty in the photometric calibration. For the weakest more extended sources marked with a $^*$, the very large quoted errors reflect the fact that the source borders are difficult to define, and hence a measurement through a circular aperture is uncertain. Still for these sources, the detection is clear, at a level of more than 6$\sigma$~above the background noise.  

($^d$): measurements performed on the deconvolved image

($^{nd}$): measurements performed on the non-deconvolved image

\end{table}
The signal-to-noise ratio of the observed PSF only allowed, among the four images obtained, the 12.8\micro~image to be deconvolved. The reduced data are presented in Fig~\ref{Fig1}: (1) 8$\times$8\arcsec$^2$~central region of the 12.8\micro~non-deconvolved image with identification of the sources, (2) 12.8\micro~non-deconvolved image (same as (1), but only the 4$\times$4\arcsec$^2$~central region), (3) 12.8\micro~deconvolved image, (4) [NeII]~continuum subtracted image, (5) 18.7\micro~image. 

The final resolution on the images can be measured by the E-W extension of the core, which is known to be unresolved down to 0.1\arcsec~(see Sect.~\ref{introduction}). The measured full widths at half maximum (FWHM) are respectively 0.29\arcsec~for images (1) and (2), 0.19\arcsec~for image (3), 0.26\arcsec~for image (4), 0.66\arcsec~for image (5).  

In a 4\arcsec~diameter aperture centred on the core, we measure the following fluxes: 43$\pm$8\,Jy in the 12.8\micro~filter and 45$\pm$9\,Jy in the 18.7\micro~filter. In the same aperture, \citet{Bock00} found 38$\pm$4\,Jy at 12.5\micro, and \citet{Tomono01} found 33.3$\pm$4\,Jy at 12.3\micro~and 46.6$\pm$4\,Jy at 18.5\micro. The larger flux value in the 12.8\micro~filter can be attributed to the presence of the [NeII] line, which contributes to about 10\% of the total flux. From the continuum subtracted [NeII] image, we find a [NeII] line flux of 2.1$\times$10$^{-12}$\,erg\,s$^{-1}$\,cm$^{-2}$ through the same 4\arcsec~diameter aperture. This flux is consistent with the ISO/SWS [NeII] 12.8\micro~flux: it represents 30\%~of 7$\times$10$^{-12}$\,erg\,s$^{-1}$\,cm$^{-2}$ measured by \citet{Lutz00} through a much larger aperture (14$\times$20\arcsec$^2$). 

On the 12.8\micro~image (Fig.~\ref{Fig1}.2), a diffraction ring can clearly be seen with a radius of about 1.2\arcsec. A residual of this ring structure remains on the continuum subtracted [NeII] image shown in Fig.~\ref{Fig1}.4. The complex morphology revealed by the VISIR images is consistent with previous observations. The image shown in Fig.~\ref{Fig1}.1 details the very extended emission already observed by \citet{Alloin00}. On the 4$\times$4\arcsec$^2$~region around the core, the deconvolved 12.8\micro~VISIR image (Fig.~\ref{Fig1}.3) is strikingly similar to the Subaru/MIRTOS 12.3\micro~map by \citet{Tomono01}, shown in Fig.~\ref{Fig1}.6. As we did not have a PSF with high enough signal-to-noise ratio to deconvolve the VISIR 18.7\micro~image, we display instead the Subaru/MIRTOS deconvolved 18.5\micro~image by \citet{Tomono01} in Fig.~\ref{Fig1}.9. 

In the top-right insert of Fig.~\ref{Fig1}.1, we present the identification of 13 sources, including the core. Sources SW3, SW4, SW5, NE4, NE5, NE6 and NE7 were identified on the non-deconvolved 12.8\micro~image, at a detection level better than  6$\sigma$ above the background noise (6$\sigma$ for source SW5). Sources SW1, NE1, NE2, NE3 were identified on the deconvolved image, and are clearly real since they can also be seen on the non-deconvolved image, and since they have been identified in previous studies \citep{Bock00,Tomono01}. Table~\ref{table} gives the positions for these sources and the 12.8\micro~filter flux densities through apertures of 0.6\arcsec~diameter (corresponding to the circles on Fig.~\ref{Fig1}).

\section{Discussion} 
\label{discussion}
In the following, we compare the MIR maps to the adaptive optics 3.5\micro~map (Fig.~\ref{Fig1}.7) from NACO \citep{Rouan04}, and to the HST/WFPC2 [OIII] $\lambda$5007 map (Fig.~\ref{Fig1}.8), obtained from the HST archive. The positioning of these maps is made according to \citet{Galliano03b}: at 3.5\micro~, the core coincides with the central engine \citep[$\equiv$ radio source S1 in ][]{Gallimore96c}, and in [OIII], the NLR cloud B is located 0.1\arcsec~north of the central engine. 

\subsection{The core}
\label{the core}
On the VISIR 12.8\micro~image, we measure the following sizes for the core: on the non-deconvolved image, the N-S and E-W FWHM are respectively 0.44\arcsec~and 0.29\arcsec, while on the deconvolved image, these values are 0.29\arcsec~and 0.19\arcsec. This is consistent with the FWHM derived by \citet{Tomono01}: 0.3\arcsec$\times$0.19\arcsec. \citet{Bock00} did not resolve the core in the E-W direction, even at 0.1\arcsec~resolution. \citet{Rouan04} give the N-S FWHM of 0.122\arcsec~for the core at 3.5\micro. Their claim to have resolved the core in the E-W direction at 2.2\micro~has not been confirmed by the K-band visibility measurements with VLTI/VINCI by \citet{Wittkowski04}, who find that the 2.2\micro~core comes from a region smaller than 5\,mas at PA$\sim$45\degr. 

The comparison of the IR maps with the [OIII] map can help understand the N-S extension of the IR core. The very good spatial coincidence between the MIR sources and the NLR [OIII] emitting clouds is well illustrated in Fig.~\ref{Fig1}. As already discussed in \citet{Galliano03b}, this suggests that the [OIII] emitting clouds are also MIR emitters. If it is so, the MIR emission in the central 0.4\arcsec~diameter region is composed of the emission from the torus itself (the core), plus the emission of the NLR cloud B, 0.1\arcsec~north, plus the (weaker) emission of NLR cloud A, 0.1\arcsec~south. The MIR emission from the core appears then naturally extended along the N-S direction. This implies that the N-S extension is not likely to be related to any intrinsic property of the dusty torus.    

\subsection{The extended emission}
\label{the extended emission}
The MIR extended emission displays an overall elongation along the NE-SW direction, similar to that of the NLR. Yet, a close look at the image reveals that the PA of the isophotes (drawn as ticks on Fig.~\ref{Fig1}.1) varies with their radius: the innermost isophote (r $\sim$ 20\,pc) is at PA=-4\degr, while the outermost one (r $\sim$ 300\,pc) is at PA=+31\degr. The twisting of the MIR emission pattern, from PA=-4\degr~to 31\degr~between radii of 20\,pc~to 300\,pc, highlights the presence of an inner spiral pattern which was already recognised on the 3.5\micro~image (Fig.~\ref{Fig1}.7). Interestingly, close to the central engine, the PA of the radio emission isophote \citep[PA=-7\degr~between the radio sources S1 and S2, S2 located to the south,][]{Gallimore96c} is very close to the PA value found on the MIR image at a scale of 20\,pc. 

The [OIII] $\lambda$5007 map (Fig~\ref{Fig1}.8) can be compared to the continuum subtracted [NeII] line image (Fig~\ref{Fig1}.4). Since this image is very similar to the 12.8\micro~deconvolved image, but has a slightly lower angular resolution, we prefer to use the 12.8\micro~filter image for the comparison. The coincidence between the NE MIR sources and the NLR clouds is very good: the MIR core includes both the NLR clouds A and B, source NE1 corresponds to the NLR clouds C and E, NE2 relates to the NLR cloud F, NE3 to the NLR cloud D and finally NE4 to the NLR cloud G. As mentioned already in \citet{Galliano03b}, this strongly supports the idea that all the MIR emission which is not related with the torus arises from dust associated with the NLR clouds. To the south, only weak [OIII] emission is found in coincidence with SW1 and SW2. 

The fact that in the [OIII] line the southern sources are weak compared to the northern ones, whereas in the [NeII] line they appear with comparable strengths, mainly occurs for an instrumental reason: the F502N filter extends, in the rest frame of NGC1068, from -1400\kms~up to +200\kms. The HST/STIS spectra presented in \citet{Cecil02} indicate that the northern clouds are moving towards us at $\sim$ 500 to 800\kms (remaining within the filter), while the southern [OIII] emission clouds are receding with velocities around 500\kms (falling outside the filter). This implies that the southern part of the NLR must actually be brighter than it appears on the HST [OIII] map. Hence, the southern MIR sources SW1 and SW2 are also likely associated with NLR region clouds. In that respect, comparison of the VISIR 12.8\micro~map with images obtained by \citet{Thompson01} in the H$\alpha$ and Pa$\alpha$ lines is enlightening. There is an overall excellent agreement between the 12.8\micro~map and the Pa$\alpha$ image: on the latter, in the SW quadrant two Pa$\alpha$ emission knots can be seen, at distances of about 2.2\arcsec~and 3.4\arcsec~from the central core. They match perfectly the source SW3 and the complex SW4 plus SW5. On the contrary, these knots are barely visible on the H$\alpha$ image. This can be interpreted as extinction of the southern cone of the NLR by a large-scale disc, roughly perpendicular to the NRL axis. The extinctions derived by \citet{Kraemer00} are $E(B-V)=0.35$ for the southern cone of the NLR, and $E(B-V)=0.22$ for the northern cone, supporting this hypothesis. 

Using the 18.5\micro~map \citep[][ Fig.~\ref{Fig1}.9]{Tomono01}, we can derive the 12.8\micro/18.5\micro~colour temperature for each aperture drawn in Fig.~\ref{Fig1}. Assuming dust black-body radiation, we find that the temperature varies from 350\,K to 200\,K as the distance from the clouds to the central engine increases, up to $\sim$4\arcsec. Hence, as this increases, the clouds become weaker in the MIR, and have cooler 18.5\micro/12.8\micro~colour, in agreement with the idea that the clouds are heated by the central engine.

\subsection{Could the MIR clouds be associated with young massive star clusters?}

The NLR clouds are generally thought to be photoionised by hard UV photons from the central engine. As we have just seen, the good correlation observed between the NIR-MIR sources and the NLR clouds supports the idea that the dominant contribution to the NIR-MIR emission comes from heated dust, surviving in the backside of dense clouds. Moreover, the spatial evolution of the MIR intensity and colours suggest heating by the central engine. Still, the question whether these clouds shelter star clusters has not been addressed yet.

The observed size of the MIR sources, of the order of 14\,pc or less, is reminiscent of the size of Young Massive Star Clusters (YMCs) which are often found in starburst regions. The measured N-band fluxes and upper limits of the L-band fluxes of the identified NLR clouds (several Jy in the N-band, and several tens of mJy in the L-band) are at least one order of magnitude larger than the typical fluxes of embedded YMCs \citep[][ and references therein]{Galliano05}. This means that the observed MIR flux comes from dust mainly heated by the AGN central engine, but does not discard the possibility for the NLR clouds to shelter YMCs. 

If the MIR sources were indeed hosting YMCs, one would expect important radio emission from the HII regions surrounding their massive stars or from their supernovae remnants. The lack of small-scale correlation between the radio and MIR emission argues against such a scenario, unless the putative YMCs have a peculiar IMF with an upper mass cutoff lower than usually considered or unless the clusters are old enough for the massive stars to have already disappeared, hence suppressing the HII region emission and the SNR radio emission. 

Let us notice finally that the observed inner spiral pattern prompts also the question of the presence of YMCs in the NLR, as starburst activity is often induced by bars or mini-bars in the central regions of galaxies. So far, the arguments do not allow to draw any clear conclusion. In order to investigate whether YMCs are sheltered by the NLR clouds, the next step is to obtain NIR-MIR spectra at high spatial and spectral resolutions of the various clouds identified in this study.  


\section*{Acknowledgements}
We acknowledge the efficient support of the Paranal Observatory Team during the SV run. We are gratefully indebted to D. Gratadour and D. Tomono for kindly providing us the NACO and MIRTOS images of NGC1068 respectively. We thank an anonymous referee for interesting comments.  


\bibliographystyle{mn2e}
\bibliography{manuref_link}

\begin{thebibliography}{}

\bibitem[\protect\citeauthoryear{{Alloin}, {Pantin}, {Lagage} \&
  {Granato}}{{Alloin} et~al.}{2000}]{Alloin00}
{Alloin} D.,  {Pantin} E.,  {Lagage} P.~O.,    {Granato} G.~L.,  2000, \aap,
  363, 926

\bibitem[\protect\citeauthoryear{{Bock}, {Marsh}, {Ressler} \& {Werner}}{{Bock}
  et~al.}{1998}]{Bock98}
{Bock} J.~J.,  {Marsh} K.~A.,  {Ressler} M.~E.,    {Werner} M.~W.,  1998,
  \apjl, 504, L5+

\bibitem[\protect\citeauthoryear{{Bock}, {Neugebauer}, {Matthews}, {Soifer},
  {Becklin}, {Ressler}, {Marsh}, {Werner}, {Egami} \& {Blandford}}{{Bock}
  et~al.}{2000}]{Bock00}
{Bock} J.~J.,  {Neugebauer} G.,  {Matthews} K.,  {Soifer} B.~T.,  {Becklin}
  E.~E.,  {Ressler} M.,  {Marsh} K.,  {Werner} M.~W.,  {Egami} E.,
  {Blandford} R.,  2000, \aj, 120, 2904

\bibitem[\protect\citeauthoryear{{Cecil}, {Dopita}, {Groves}, {Wilson},
  {Ferruit}, {P{\' e}contal} \& {Binette}}{{Cecil} et~al.}{2002}]{Cecil02}
{Cecil} G.,  {Dopita} M.~A.,  {Groves} B.,  {Wilson} A.~S.,  {Ferruit} P.,
  {P{\' e}contal} E.,    {Binette} L.,  2002, \apj, 568, 627

\bibitem[\protect\citeauthoryear{{Cohen}, {Walker}, {Carter}, {Hammersley},
  {Kidger} \& {Noguchi}}{{Cohen} et~al.}{1999}]{Cohen99}
{Cohen} M.,  {Walker} R.~G.,  {Carter} B.,  {Hammersley} P.,  {Kidger} M.,
  {Noguchi} K.,  1999, \aj, 117, 1864

\bibitem[\protect\citeauthoryear{{Efstathiou} \& {Rowan-Robinson}}{{Efstathiou}
  \& {Rowan-Robinson}}{1995}]{Efstathiou95}
{Efstathiou} A.,  {Rowan-Robinson} M.,  1995, \mnras, 273, 649

\bibitem[\protect\citeauthoryear{{Evans}, {Ford}, {Kinney}, {Antonucci},
  {Armus} \& {Caganoff}}{{Evans} et~al.}{1991}]{Evans91}
{Evans} I.~N.,  {Ford} H.~C.,  {Kinney} A.~L.,  {Antonucci} R.~R.~J.,  {Armus}
  L.,    {Caganoff} S.,  1991, \apjl, 369, L27

\bibitem[\protect\citeauthoryear{{Galliano}, {Alloin}, {Granato} \&
  {Villar-Mart{\'{\i}}n}}{{Galliano} et~al.}{2003}]{Galliano03b}
{Galliano} E.,  {Alloin} D.,  {Granato} G.~L.,    {Villar-Mart{\'{\i}}n} M.,
  2003, \aap, 412, 615

\bibitem[\protect\citeauthoryear{{Galliano}, {Alloin}, {Pantin}, {Lagage} \&
  {Marco}}{{Galliano} et~al.}{2005}]{Galliano05}
{Galliano} E.,  {Alloin} D.,  {Pantin} E.,  {Lagage} P.,    {Marco} O.,  2005,
  astro-ph/0504118

\bibitem[\protect\citeauthoryear{{Gallimore}, {Baum} \& {O'Dea}}{{Gallimore}
  et~al.}{1996}]{Gallimore96c}
{Gallimore} J.~F.,  {Baum} S.~A.,    {O'Dea} C.~P.,  1996, \apj, 464, 198

\bibitem[\protect\citeauthoryear{{Granato} \& {Danese}}{{Granato} \&
  {Danese}}{1994}]{Granato94}
{Granato} G.~L.,  {Danese} L.,  1994, \mnras, 268, 235+

\bibitem[\protect\citeauthoryear{{Granato}, {Danese} \&
  {Franceschini}}{{Granato} et~al.}{1997}]{Granato97}
{Granato} G.~L.,  {Danese} L.,    {Franceschini} A.,  1997, \apj, 486, 147+

\bibitem[\protect\citeauthoryear{{Gratadour}, {Rouan}, {Boccaletti}, {Riaud} \&
  {Cl{\' e}net}}{{Gratadour} et~al.}{2005}]{Gratadour05}
{Gratadour} D.,  {Rouan} D.,  {Boccaletti} A.,  {Riaud} P.,    {Cl{\' e}net}
  Y.,  2005, \aap, 429, 433

\bibitem[\protect\citeauthoryear{{Jaffe}, {Meisenheimer}, {R{\" o}ttgering},
  {Leinert}, {Richichi}, {Chesneau}, {Fraix-Burnet}, {Glazenborg-Kluttig},
  {Granato}, {Graser} \& {Heijligers}}{{Jaffe} et~al.}{2004}]{Jaffe04}
{Jaffe} W.,  {Meisenheimer} K.,  {R{\" o}ttgering} H.~J.~A.,  {Leinert} C.,
  {Richichi} A.,  {Chesneau} O.,  {Fraix-Burnet} D.,  {Glazenborg-Kluttig} A.,
  {Granato} G.-L.,  {Graser} U.,    {Heijligers} 2004, \nat, 429, 47

\bibitem[\protect\citeauthoryear{{Kraemer} \& {Crenshaw}}{{Kraemer} \&
  {Crenshaw}}{2000}]{Kraemer00}
{Kraemer} S.~B.,  {Crenshaw} D.~M.,  2000, \apj, 544, 763

\bibitem[\protect\citeauthoryear{{Krolik} \& {Begelman}}{{Krolik} \&
  {Begelman}}{1986}]{Krolik86}
{Krolik} J.~H.,  {Begelman} M.~C.,  1986, \apjl, 308, L55

\bibitem[\protect\citeauthoryear{{Lutz}, {Sturm}, {Genzel}, {Moorwood},
  {Alexander}, {Netzer} \& {Sternberg}}{{Lutz} et~al.}{2000}]{Lutz00}
{Lutz} D.,  {Sturm} E.,  {Genzel} R.,  {Moorwood} A.~F.~M.,  {Alexander} T.,
  {Netzer} H.,    {Sternberg} A.,  2000, \apj, 536, 697

\bibitem[\protect\citeauthoryear{{Macchetto}, {Capetti}, {Sparks}, {Axon} \&
  {Boksenberg}}{{Macchetto} et~al.}{1994}]{Macchetto94}
{Macchetto} F.,  {Capetti} A.,  {Sparks} W.~B.,  {Axon} D.~J.,    {Boksenberg}
  A.,  1994, \apjl, 435, L15

\bibitem[\protect\citeauthoryear{{Marco} \& {Alloin}}{{Marco} \&
  {Alloin}}{2000}]{Marco00}
{Marco} O.,  {Alloin} D.,  2000, \aap, 353, 465

\bibitem[\protect\citeauthoryear{{Pier} \& {Krolik}}{{Pier} \&
  {Krolik}}{1993}]{Pier93}
{Pier} E.~A.,  {Krolik} J.~H.,  1993, \apj, 418, 673+

\bibitem[\protect\citeauthoryear{{Rouan}, {Lacombe}, {Gendron}, {Gratadour},
  {Cl{\' e}net}, {Lagrange}, {Mouillet}, {Boisson}, {Rousset}, {Fusco},
  {Mugnier}, {S{\' e}chaud}, {Thatte}, {Genzel}, {Gigan}, {Arsenault} \&
  {Kern}}{{Rouan} et~al.}{2004}]{Rouan04}
{Rouan} D.,  {Lacombe} F.,  {Gendron} E.,  {Gratadour} D.,  {Cl{\' e}net} Y.,
  {Lagrange} A.-M.,  {Mouillet} D.,  {Boisson} C.,  {Rousset} G.,  {Fusco} T.,
  {Mugnier} L.,  {S{\' e}chaud} M.,  {Thatte} N.,  {Genzel} R.,  {Gigan} P.,
  {Arsenault} R.,    {Kern} P.,  2004, \aap, 417, L1

\bibitem[\protect\citeauthoryear{{Rouan}, {Rigaut}, {Alloin}, {Doyon}, {Lai},
  {Crampton}, {Gendron} \& {Arsenault}}{{Rouan} et~al.}{1998}]{Rouan98}
{Rouan} D.,  {Rigaut} F.,  {Alloin} D.,  {Doyon} R.,  {Lai} O.,  {Crampton} D.,
   {Gendron} E.,    {Arsenault} R.,  1998, \aap, 339, 687

\bibitem[\protect\citeauthoryear{{Starck}, {Pantin} \& {Murtagh}}{{Starck}
  et~al.}{2002}]{Starck02}
{Starck} J.~L.,  {Pantin} E.,    {Murtagh} F.,  2002, \pasp, 114, 1051

\bibitem[\protect\citeauthoryear{{Thompson}, {Chary}, {Corbin} \&
  {Epps}}{{Thompson} et~al.}{2001}]{Thompson01}
{Thompson} R.~I.,  {Chary} R.,  {Corbin} M.~R.,    {Epps} H.,  2001, \apjl,
  558, L97

\bibitem[\protect\citeauthoryear{{Tomono}, {Doi}, {Usuda} \&
  {Nishimura}}{{Tomono} et~al.}{2001}]{Tomono01}
{Tomono} D.,  {Doi} Y.,  {Usuda} T.,    {Nishimura} T.,  2001, \apj, 557, 637

\bibitem[\protect\citeauthoryear{{Wittkowski}, {Kervella}, {Arsenault},
  {Paresce}, {Beckert} \& {Weigelt}}{{Wittkowski} et~al.}{2004}]{Wittkowski04}
{Wittkowski} M.,  {Kervella} P.,  {Arsenault} R.,  {Paresce} F.,  {Beckert} T.,
     {Weigelt} G.,  2004, \aap, 418, L39

\end{thebibliography}
\bsp
\label{lastpage}
\end{document}